%% file: main_text_with_figs.tex
\documentclass[%
 letterpaper,twocolumn,english,
 superscriptaddress,
showpacs,preprintnumbers,
 amsmath,amssymb,
 aps,
 prc,
floatfix,
]{revtex4-1}

\usepackage[usenames]{color}
\usepackage{amsmath}
\usepackage{amssymb}
\usepackage{graphicx}% Include figure files
\usepackage[dvipdfm,ps2pdf,unicode=true,bookmarks=false,breaklinks,pdfborder={0
    0 0},backref=false,colorlinks=true]{hyperref}  
\hypersetup{pdftitle={Single Spin Asymmetries of Inclusive Hadrons
    Produced in Electron Scattering from a Transversely Polarized
    He-3 Target},pdfauthor={K. Allada, et. al. (The Jefferson Lab Hall
    A Collaboration)},pdfsubject={v6.4: nucl-ex,
    hep-ex},pdfkeywords={Transverse SSA Inclusive hadron SIDIS He-3
    AN TMD},linkcolor=DarkBlueCite, citecolor=DarkBlueCite,
  urlcolor=DarkBlueCite}  
\usepackage[hyphenbreaks]{breakurl}

\makeatletter

\usepackage{dcolumn}% Align table columns on decimal point
\usepackage{bm}% bold math

\definecolor{DarkBlueCite}{rgb}{0.1,0.0,0.5}

\makeatother

\begin{document}
\input{author_list.tex}

%\preprint{APS/123-QED}
\title{Single Spin Asymmetries of Inclusive Hadrons Produced in
  Electron Scattering from a Transversely Polarized $^3$He
  Target} 

\date{\today}
%\vspace*{0.1cm}

\begin{abstract}
We report the first measurement of target single-spin asymmetries (A$_N$)
in the inclusive hadron production reaction,
$e~$+$~^3\text{He}^{\uparrow}\rightarrow h+X$, using a transversely polarized
$^3$He target. The experiment was conducted at Jefferson Lab in Hall A using a
5.9-GeV electron beam. Three types of hadrons ($\pi^{\pm}$,
$\text{K}^{\pm}$ and proton) were detected in the transvere hadron
momentum range \mbox{0.54 $<p_T<$ 0.74 GeV/c}. The range of $x_F$ for
pions was -0.29 $<x_F<$ -0.23 and for kaons \mbox{-0.25 $<x_F<$
  -0.18}. The observed asymmetry strongly depends on the type of
hadron. A positive   asymmetry is observed for $\pi^+$ and
$\text{K}^+$. A negative asymmetry is observed for $\pi^{-}$. The
magnitudes of the asymmetries follow $|A^{\pi^-}| < |A^{\pi^+}| <
|A^{K^+}|$. The K$^{-}$ and proton asymmetries are consistent with
zero within the experimental uncertainties. The $\pi^{+}$ and
$\pi^{-}$ asymmetries measured for the $^3$He target and extracted
for neutrons are opposite in sign with a small increase observed as a
function of $p_T$.
\end{abstract}


\maketitle
%\vspace{5.0cm}

%xxx.xx Introduction. xxx.xx
The study of the transverse single spin asymmetries (TSSAs) 
is one of the most active areas of research in modern hadronic physics. 
TSSA is an important tool to advance our understanding of 
the nucleon spin, to reveal the role of the quark orbital angular
momentum (OAM), and to access the three-dimensional structure of the
nucleon in momentum space~\cite{Barone2010}. Current research
on TSSA focuses on the polarized proton-proton ($pp^{\uparrow}$)
and lepton-nucleon ($lN^{\uparrow}$) reaction channels. 

An early observation of large left-right SSAs (A$_N$) in the
$pp^{\uparrow}\rightarrow\pi^{\pm}X$ 
reaction by the Fermilab E704 experiment at $\sqrt{s}$=19.4
GeV~\cite{E704_1,E704_2} revealed a strong dependence on the hadron
type. In the center-of-mass frame of the polarized
$pp^{\uparrow}$ collision, viewed along the momentum direction of the
polarized proton, $\pi^{+}$ favors the left side of the 
spin vector, whereas $\pi^{-}$ favors the right side of the
spin vector. More recently, such non-vanishing TSSAs were
observed for $\pi^{\pm}$ and K$^{\pm}$ at $\sqrt{s}$=62.4 GeV by
BRAHMS~\cite{BRAHMS2008}, and for neutral pions at $\sqrt{s}$=200 GeV
by the STAR experiment at RHIC~\cite{STAR2008}. Although TSSAs
have been observed in $pp^{\uparrow}$ reactions for more than two
decades, measurement in semi-inclusive deep inelastic
scattering (SIDIS) is regarded as one of the cleanest ways to
understand them at the partonic level. TSSAs have been measured in the
SIDIS reaction ($lp^{\uparrow}\rightarrow l'hX$) by
HERMES~\cite{HERMES2005,HERMES2009,HERMES2010,HERMES2013} with a polarized proton
target, and by
COMPASS~\cite{COMPASS2010,COMPASS2009,COMPASS2012,COMPASS2012_1} using
polarized proton and deuteron targets. Recently, they have been measured at
Jefferson Lab in Hall A using a polarized $^3$He
target~\cite{Qian2011,Huang2012}. 

The origin of TSSAs is currently interpreted using two theoretical 
approaches~\cite{Anselmino2010}. The first approach
is based on the transverse-momentum-dependent distribution 
and fragmentation functions (TMDs) in the framework of the TMD
factorization, and is mostly used to explain TSSAs in the SIDIS process. 
There are two reaction mechanisms: the Collins effect~\cite{Collins1993}
and the Sivers effect~\cite{Sivers1990}. In the Collins effect, 
the TSSA is generated by the transversity distribution, which represents 
the probability of finding a transversely polarized parton inside a
transversely polarized nucleon, and the Collins fragmentation function, 
which correlates the transverse polarization of the quark with the transverse 
momentum of the outgoing hadron ($p_T$). In the Sivers effect, the
TSSA is generated by the Sivers distribution function, which
correlates the quark's transverse momentum and the nucleon's spin, and is
sensitive to the quark OAM. More specifically, the observed asymmetry due
to the Sivers function arises from the final-state interaction between the
struck quark and the nucleon remnant in SIDIS. On the other hand, the
Sivers function in the Drell-Yan process is expected to arise from the
initial-state interactions~\cite{stan2002}. Taking gauge links into
consideration, the Sivers distribution is predicted to be process
dependent in the sense that it differs in sign between SIDIS and
Drell-Yan processes~\cite{Collins200243,stan2002}. Furthermore, in
models such as the di-quark model~\cite{Lu-Schmidt2007}, one can
connect the Sivers distribution for each quark to its contribution to
the anomalous magnetic moment of the nucleon.

The second approach is based on the twist-3 collinear
factorization~\cite{Efremov1982,Efremov1985,Qiu1991,Qiu1998}, where
the SSAs are interpreted in terms of higher-twist quark-gluon
correlations, and is mainly used to explain the TSSAs in the
$pp^{\uparrow} \rightarrow hX$ channel. It was also shown that the TMD
factorization and twist-3 methods are
related~\cite{Ji2006,Boer2003,Yuan2009}. However, the 
Sivers function extracted from $pp^{\uparrow}$ data with the twist-3
approach is shown to have a ``sign mismatch'' when compared to the
Sivers function extracted from SIDIS data. The sign mismatch indicates
a potential inconsistency in the current theoretical
formalism~\cite{Kang2012}, and needs to be further investigated. In
order to understand the underlying mechanism producing TSSA it is
crucial to study additional reaction channels~\cite{Anselmino2010,Kang2011}.

In this letter, we study TSSA from one of the experimentally least explored
reactions, inclusive hadron production using a lepton beam on a
transversely polarized nucleon ($lN^{\uparrow}\rightarrow
hX$)~\cite{Anselmino2000,Anselmino2010}. An early study of this
process was done by Anselmino {\it et al.,} under the assumption that
the underlying mechanism that generates TSSA is either Collins or
Sivers effect~\cite{Anselmino2000}. 
More recently, this study was re-evaluated using newly available SIDIS data on
Sivers and Collins moments assuming that the TMD
factorization is valid in $lp^{\uparrow}\rightarrow\pi^{\pm}X$ processes
at large $p_T$  values. Due to the presence of only one hard scale in this
process, estimation of the asymmetries is generally done at large
$p_T$ values (typically $>$1~GeV/c). They predicted asymmetries between
5-10\% for $\sqrt{s}\simeq$4.9 GeV, $p_T$=1.5 GeV/c, and $x_F\leq$0.1
with a contribution from the Sivers mechanism to A$_N$, whereas the
contribution from the Collins mechanism was negligible~\cite{Anselmino2010}.

Non-zero SSAs were also estimated based on the twist-3 distribution and
fragmentation functions in the framework of collinear
factorization~\cite{Koike2003aip,Koike2003, Kang2011}, and in the
SIDIS process by integrating over the scattered electron's azimuthal
angle~\cite{Sun2010}. These studies of TSSAs in the
$lp^{\uparrow}\rightarrow\pi^{\pm}X$ process are performed under the
assumption of a SIDIS reaction, in which hard scattering occurs
between a virtual photon and a quark. However, since the process is
dominated by the cross-section at Q$^2$ near zero, it was also pointed out
that the $lp^{\uparrow}\rightarrow\pi^{\pm}X$ process will have significant
contributions from soft processes, such as vector meson dominance,
especially at lower $p_T$ values~\cite{Afanasev2000}. Experimental
data for TSSA in this process have recently been reported by the
HERMES Collaboration using $e^-/e^+$ beams on a transversely polarized
hydrogen target~\cite{HERMES2013}.
 
\begin{figure}[ht]
\includegraphics[width=80mm]{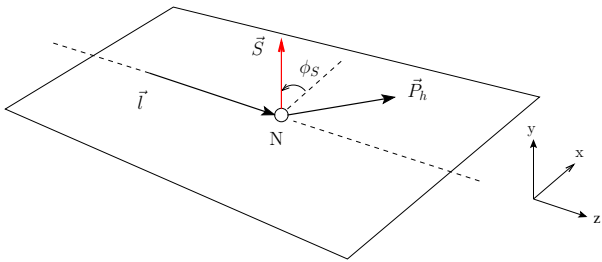}
\caption{\label{fig:fig1} (Color online) Kinematical configuration in the laboratory
  coordinate system for the $lN^{\uparrow}\rightarrow hX$
  process. $\vec{P}_h$ represents the momentum direction of the
  produced hadron, and $\vec{S}$ is the spin vector of the
  nucleon. The polarized nucleon's momentum is along the -z direction
  in the e+N center-of-mass frame} 
\end{figure}

We report the first measurement of target single-spin
asymmetries (A$_N$) in inclusive hadron ($\pi^{\pm}$, K$^{\pm}$, and
proton) production at fixed-target e+N center-of-mass energy
$\sqrt{s}=$3.45 GeV, using an unpolarized electron beam and a
transversely polarized $^3$He target as an effective polarized neutron
target. The kinematical variables for this process are: $x_F =
2p^{CM}/\sqrt{s}$, where $p^{CM}$ is the momentum of the outgoing
hadron along the polarized nucleon's momentum direction in the e+N
center-of-mass frame, and $p_T=\sqrt{p_x^2+p_y^2}$, the transverse
momentum of the outgoing hadron. The kinematical configuration in the
laboratory coordinate system is shown in Fig~\ref{fig:fig1}. 
The target spin ``up''($\uparrow$) was defined to be along the +$\hat{y}$
direction, parallel to the vector $\vec{l}\times\vec{P_h}(\phi_S =
90^{\circ})$, where $\vec{l}$ and $\vec{P_h}$ are the momentum vectors
of the incoming beam and outgoing hadron, respectively. 

The target SSA is defined as~\cite{Anselmino2010},
\begin{equation}
A_{UT}(x_F,p_T) =
\frac{1}{P}\frac{d\sigma^{\uparrow}-d\sigma^{\downarrow}}{d\sigma^{\uparrow}+d\sigma^{\downarrow}}\text{sin}\phi_S
= A_N\text{sin}\phi_S,
\label{eqn:ssa_def}
\end{equation}
where $d\sigma^{\uparrow(\downarrow)}$ is the differential cross-section
in the target ``up''(``down'') state, and $P$ is the target
polarization. The spin-dependent part of the cross-section is
proportional to the term
$\vec{S}\cdot(\vec{l}\times\vec{P_h}$), which gives rise to
a $\text{sin}(\phi_S)$ modulation in the definition of the asymmetry. This
term makes A$_N$ parity-conserving, but T-odd under ``na\"{\i}ve''
time reversal, in which the initial and final states do not
interchange. Note that the sign of $A_N$ in the laboratory coordinate system
of this experiment (Eq. 1) differ by a factor -1
from the definition in the phenomenological study of this process
in~\cite{Anselmino2010}, where the authors used the center-of-mass
coordinate system with the lepton moving in the -$\hat{z}$ direction. 

The data were collected using a singles trigger during the E06-010
experiment in Hall A at Jefferson Lab~\cite{allada2010}. A beam
energy of 5.9 GeV was provided by the CEBAF accelerator with an average
current of 12 $\mu A$. The produced hadrons were detected in a
high-resolution spectrometer (HRS)~\cite{Alcorn2004} at a central
angle of 16$^{\circ}$ on beam left side. Positively and negatively charged
particles were detected separately by changing the magnet polarity of
the HRS. The central momentum of the HRS was fixed at 2.35~GeV/c, with
a momentum acceptance of $\pm$4.5$\%$ and solid angle of $~$6
msr. The average transverse momentum of the detected hadrons
($<$$p_T$$>$) was 0.64  GeV/c. We note that if the pions are
produced through virtual or real-photon exchange ($\gamma^\star$+N or
$\gamma$+N), the minimum photon energy is E$_{\gamma}\ge$2.6
GeV, corresponding to an invariant mass of W$\geq$2.4 GeV for the
$\gamma$+N system, well above the region of nucleon resonances.

The data from the two helicity states from the polarized electron beam
were summed over to achieve an unpolarized beam. The residual
helicity-sorted beam-charge difference was less than 100 ppm in a
typical run. The target spin direction was automatically reversed ($\phi_S=\pm
90^{\circ}$) at a rate of once every 20 minutes, which allowed control
of the combined systematic uncertainty due to luminosity fluctuations 
and time depedence to below 50 ppm in this experiment.

Polarized $^3$He targets have often been used as an effective
polarized neutron targets, because in the ground state of the $^3$He
nuclear wavefunction (dominated by the S-state) the two proton spins are
opposite to each other, and the nuclear spin is carried by the remaining
neutron~\cite{Bissey2001}. The polarized target used in this
measurement was a 40-cm long glass cell filled with $\sim$8 atm of
$^3$He gas and a small amount ($\sim$0.13 atm) of N$_2$ gas to reduce
depolarizing effects~\cite{Alcorn2004}. The radiation lengths of the
materials up to the center of the $^3$He target were: Be window
(0.072\%), $^4$He gas (0.004\%), glass window (0.142\%), and $^3$He gas
(0.046\%). The target was polarized via hybrid spin-exchange optical
pumping of a mixture of Rb-K~\cite{Babcock2005}. The $^3$He polarization
was measured every 20 minutes during the spin-reversal
using Nuclear Magnetic Resonance (NMR). The NMR signal was calibrated
with Electron Paramagnetic Resonance (EPR) measurements and a known
NMR signal obtained from an identical water cell. The average in-beam
polarization of the target was (55.4$\pm$2.8)$\%$.   

The HRS detector package consisted of four separate
detectors for particle identification: (i) a light-gas threshold
\v{C}erenkov for electron identification, (ii) a two\nobreakdash-layer
electromagnetic calorimeter for electron-hadron separation,
(iii) a threshold aerogel \v{C}erenkov detector for pion
identification, and (iv) a ring imaging \v{C}erenkov (RICH) detector
for $\pi^{\pm}$, K$^{\pm}$, and proton identification. The electron
and positron background were suppressed with a rejection factor of
10$^4$:1. After all the particle identification cuts the contamination due
to leptons was negligible in the hadron sample. The pion sample
had a contamination of $<$1\% due to other hadrons. Kaons were
identified using the RICH detector, in combination
with a veto from the aerogel counter, to suppress the large pion
background. To further improve the purity of the kaon sample, a
$\chi^2$ probability distribution was constructed based on the
reconstructed \v{C}erenkov ring angle and the expected \v{C}erenkov angle
in the RICH detector for a known particle
momentum~\cite{Urciuoli2009}. A cut on this distribution effectively
suppresses the background events due to mis-identified particles. The
contamination of the kaon sample from other hadrons was estimated to
be $\sim 3\%$ (proton) and $\sim 2\%$ ($\pi^+$) for positive kaons, and
$\sim 2\%$ ($\pi^-$) for negative kaons. Protons were identified using
the same method that was used for charged kaons, producing a very clean
sample with estimated background $<$1\%.

The raw $^3$He target single-spin asymmetry ($A_N$) was obtained using the
normalized yields in target spin up/down ($\phi_S= \pm
90^{\circ}$) states, as shown in Eq.~\ref{eqn:ssa_def}. The yield in each
spin state is normalized with the accumulated beam charge and livetime
of the data acquisition system in that state. The dilution of the
measured $^3$He asymmetries due to the presence of a small amount of
N$_2$ gas in the target cell was corrected using the
factor, 
\begin{equation}
f_{\text{N}_{2}}\equiv\frac{\rho_{\text{N}_{2}}\sigma_{\text{N}_{2}}}{\rho_{^{3}\text{He}}\sigma_{^{3}\text{He}}+\rho_{\text{N}_{2}}\sigma_{\text{N}_{2}}},
\label{eqn:N2dil}
\end{equation}
where $\rho$ is the density of the gas in the cell and $\sigma$ is the
unpolarized inclusive hadron cross-section. The unpolarized N$_2$ and $^3$He
cross-sections were obtained from the data taken during the experiment
using reference cells filled with pure N$_2$ and $^3$He gas. The
$f_{N2}$ was extracted separately for all hadrons and was about
$\sim$10\% in each case.

The overall systematic uncertainty in this measurement was small due
to frequent target spin flips. The false asymmetry
due to luminosity fluctuations was less than 0.04$\%$ and was
confirmed by measuring the target SSA in inclusive $(e,e')$ DIS
reaction for in-plane transverse target ($\phi_S$ = 0$^{\circ}$,
180$^{\circ}$). This configuration was
achieved by rotating the target spin by 90$^{\circ}$ while keeping all
other conditions the same. This type of asymmetry vanishes under
parity conservation, assuming one-photon exchange. In addition, the
inclusive pion asymmetry was measured to be zero with a precision of
0.05\% in the same configuration ($\phi_S$ =
0$^{\circ}$, 180$^{\circ}$). This asymmetry is expected to vanishes due
to sin($\phi_S$) moment.

There were two additional sources of systematic uncertainty
associated with the RICH detector for kaons and protons. The first one
was from a cut on the number of hits in the RICH detector. The
relative change in asymmetry under variation of the cut threshold was
assigned as a systematic uncertainty. For K$^{\pm}$ it was $<$14\% and
for protons it was $<$3\%, relative to the statistical
uncertainty. The second source was local fluctuations in the kaon and
proton yield arising from detector inefficiencies in certain periods
of the data-taking. The systematic uncertainty was estimated using
the change in the asymmetry obtained in the periods with and without
these fluctuations, and was estimated to be $<$2\%, $<$6\%, and
$<$1\%, relative to the statistical uncertainty, for K$^+$, K$^-$ and
protons, respectively. Systematic uncertainties due to the
target density fluctuations, vertex cuts, DAQ livetime, and HRS
single-track events were negligible.  
\begin{figure}[ht]
\includegraphics[width=65mm]{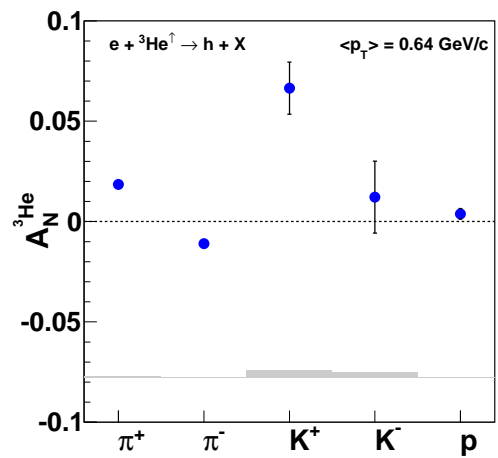}
\caption{\label{fig:fig2} (Color online) Inclusive SSA results on a $^3$He target
  for $\pi^{\pm}$, K$^{\pm}$ and protons in the vertical target spin
  configuration ($\phi_S= \pm 90^{\circ}$). The error bars on the
  points represents the statistical uncertainty. The grey band shows
  the magnitude of the overall systematic uncertainty for each hadron
  channel.}
\end{figure}

\begin{figure}[ht]
\includegraphics[width=60mm]{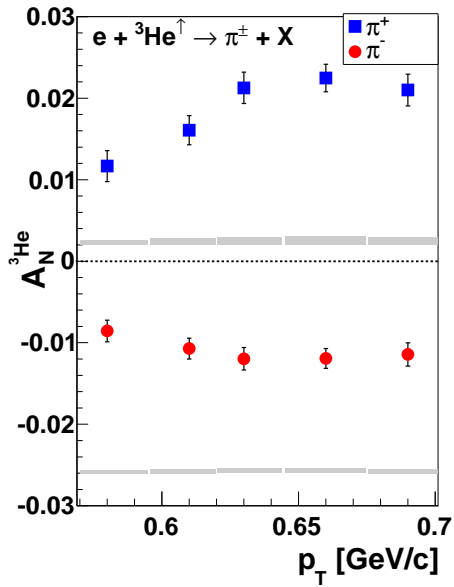}
\caption{\label{fig:fig3} (Color online) $A_N$ results on a $^3$He target
  for the $\pi^{\pm}$ channel as a function of $p_T$. The solid band
  on the bottom of each panel shows the magnitude of the systematic
  uncertainty for each momentum bin.}
\end{figure}

The final $^3$He asymmetry results are shown for different hadron
species in Fig.~\ref{fig:fig2}. These results include a small
correction due to particle contamination for each hadron species. In
Fig.~\ref{fig:fig2} the data were integrated in $p_T$ and $x_F$
(see Table~\ref{table:kine}). The error bars represents the statistical
uncertainty. The systematic uncertainties are shown as a solid
band. The measured A$_N$ for $\pi^+$($\sim$2$\%$) and
K$^+$($\sim$6$\%$) are positive, and opposite in sign to that of
$\pi^-$ ($\sim$1$\%$). In addition,
the magnitudes of these asymmetries follow $|A^{\pi^-}| < |A^{\pi^+}| < |A^{K^+}|$.
The measured A$_N$ for K$^-$ and protons was found to be small and 
consistent with zero. We note that the majority of the detected
protons originate through a knock-out reaction from the $^3$He nucleus,
whereas mesons are produced either through fragmentation process or
in a photoproduction reaction. The SSAs for charged pions as a function of
$p_T$ for a $^3$He target are shown in Fig.~\ref{fig:fig3}. The
asymmetry grows as a function of $p_T$ and plateaus around $p_T\simeq
0.63$ GeV/c. 
\begin{table}[h]
%\centering
\begin{tabular}{c|c|c|c}\hline \hline
Hadron&~~~$<$$x_F$$>$~&$<$$p_T$$>$~&~$A^{3\text{He}}_N\pm$ Stat.$\pm$ Sys.~\\
& &(GeV/c) &  \\ \hline
$\pi^{+}$ &~-0.262~ & 0.64~ &0.0185$\pm$0.0007$\pm$0.0009  \\ 
$\pi^{-}$ &~-0.262~ & 0.64~ &-0.0109$\pm$0.0005$\pm$0.0005 \\ 
K$^{+}$ &~-0.215~ & 0.64~ &0.0665$\pm$0.0130$\pm$0.0038 \\ 
K$^{-}$ &~-0.215~ & 0.64~ &0.0122$\pm$0.0179$\pm$0.0027 \\ 
$p$ &~-0.087~ & 0.64~ & 0.0038$\pm$0.0026$\pm$0.0002 \\\hline
\hline
\end{tabular}
%\centering
  \caption{Central kinematics for three types of hadrons along with
    the A$_N$ results for a $^3$He target. A negative $x_F$ indicates
    that the produced hadron is moving backwards with respect to the
    nucleon momentum direction in the center-of-mass frame of the e+N
    system.}
\label{table:kine}
\end{table}

We extracted $A_N$ on neutron from the
measured $^3$He asymmetry using the effective polarization
approach, previously used for both inclusive and
semi-inclusive DIS processes~\cite{scopetta2007,Bissey2001}. Using
this method, $A_N$ for the neutron can be obtained from
$^3$He results using the relation,
\begin{equation}
\text{A}^{^{3}\text{He}}_{\text{N}} =  \text{P}_n(1-f_p)\text{A}^{n}_{\text{N}} + \text{P}_pf_p\text{A}^p_{\text{N}},
\label{eqn:neutron_ext}
\end{equation}
where $\text{A}^{^{3}\text{He}}_{\text{N}}$ is the measured $^3$He asymmetry. P$_n$ =
0.86$^{+0.036}_{-0.02}$ and P$_p$ = -0.028$^{+0.009}_{-0.004}$ are the
effective polarization of the neutron and proton, respectively. Hence, the
contribution of proton polarization ($\simeq 2.8\%$) to
$\text{A}^{^{3}\text{He}}_{\text{N}}$ is relatively small. The factor,
$f_p$ = $\frac{2\sigma_p}{\sigma_{^{3}He}}$, in $^3$He was measured 
directly in this experiment using the 
yields obtained from unpolarized hydrogen and $^3$He targets. The
average proton dilution (1-$f_p$) for $\pi^+$ was 0.156$\pm$0.007 and
for $\pi^-$ it was 0.268$\pm$0.005. The SSA from a polarized proton
target (A$^p_{N}$) was assumed to be no more than $\pm$5\% at
$p_T\simeq$ 0.64 GeV/c, which is consistent with the
HERMES data on $A^p_N$~\cite{HERMES2013}.

\begin{figure}[ht]
\includegraphics[width=60mm]{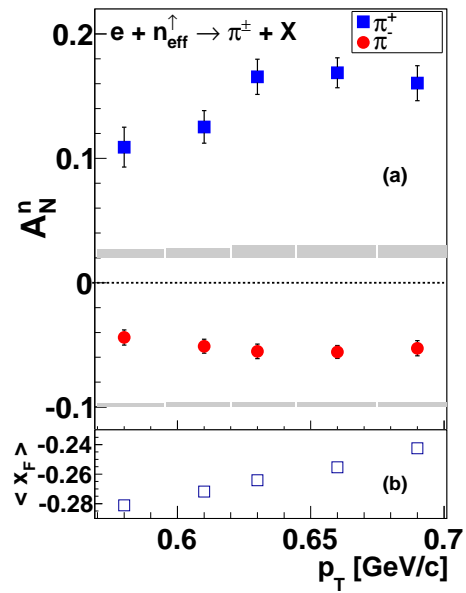}
\caption{\label{fig:fig4}(Color online) (a) A$_N$ results
  on a neutron target extracted from the measured $^3$He
  asymmetries. The solid band on the bottom of each panel shows the
  magnitude of the systematic uncertainty for each momentum bin. The lower
  plot (b) is the $x_F$ and $p_T$ correlation in this measurement.}
\end{figure}

The final results for $A^n_N$ for charged pions on an effective
neutron target are shown in Fig.~\ref{fig:fig4}. The extracted
$A^n_N$ is below 20\% for both $\pi^+$ and  $\pi^-$, with the asymmtry
amplitude for $\pi^+$ being larger than those for $\pi^-$. The $A^n_N$
for both $\pi^+$ and $\pi^-$ increase up to $p_T\simeq$0.63 GeV/c,
before it plateaus. Currently there are no theoretical estimates for A$_N$ at
$\sqrt{s}=$ 3.45 GeV and $<$$p_T$$>\sim$0.64 GeV/c for a neutron
target. The existing predictions were done for a proton target at $p_T=$
1.5 GeV/c and $\sqrt{s}\simeq 4.9~\text{GeV}$~\cite{Anselmino2010}. 
However, the sign of A$_N$ for $\pi^{\pm}$ in our experiment 
is consistent with the existing predictions dominated by the Sivers
effect, assuming p$\leftrightarrow$n and $\pi^+\leftrightarrow\pi^-$.

\begin{table}[ht]
%\centering
\begin{tabular}{c|c|c|c}\hline \hline
$<$$p_T$$>$&$A^n_N (\pi^+)\pm$Stat.$\pm$Sys.&$A^n_N
  (\pi^-)\pm$Stat.$\pm$Sys.&R$_{\pi^+/\pi^-}$\\
(GeV/c)& & &  \\
\hline
0.58&0.109$\pm$0.016$\pm$0.007&-0.044$\pm$0.006$\pm$0.003&-2.5$\pm$0.5 \\ 
0.61&0.125$\pm$0.013$\pm$0.008&-0.051$\pm$0.005$\pm$0.003&-2.5$\pm$0.4\\ 
0.63&0.166$\pm$0.014$\pm$0.010&-0.055$\pm$0.006$\pm$0.004&-3.0$\pm$0.5\\ 
0.66&0.169$\pm$0.012$\pm$0.010&-0.056$\pm$0.005$\pm$0.004&-3.0$\pm$0.4\\ 
0.69&0.160$\pm$0.014$\pm$0.010&-0.053$\pm$0.006$\pm$0.003&-3.0$\pm$0.5\\\hline
\hline
\end{tabular}
%\centering
  \caption{The extracted neutron A$^n_N$ results for $\pi^+$ and
    $\pi^-$ along with their ratio R$_{\pi^+/\pi^-}$ in five different
    $<$$p_T$$>$ bins.}
\label{table:neutron_results}
\end{table}

We can compare the observed behavior of our data
with existing TSSAs in both proton-proton ($pp^{\uparrow}$) and
lepton-nucleon ($lN^{\uparrow}$) reaction channels. Our results show
that in the center-of-mass frame of the polarized neutron-electron
collision, viewed along the direction of the neutron's momentum,
$\pi^+$ favors the right side of the spin vector, whereas $\pi^-$
favors the left side of the spin vector. Assuming isospin symmetry,
this behavior is the same as that observed in $pp^{\uparrow}
\rightarrow hX$ for the E704~\cite{E704_1,E704_2} and
BRAHMS~\cite{BRAHMS2008} experiments. In addition,
  this behavior is also the same as the Collins asymmetry for
  $\pi^{\pm}$, and the Sivers asymmetry for $\pi^+$ observed in
SIDIS~\cite{HERMES2005,HERMES2009,HERMES2010,COMPASS2012,COMPASS2012_1,Qian2011}.  
The A$^n_N$  for $\pi^+$ is about $\sim$ 15\% at $<$$p_T$$>$=0.64 GeV/c,
which is larger than that for HERMES proton data for $\pi^+$ ($\sim$ 5\% at
$<$$p_T$$>$=0.68 GeV/c)~\cite{HERMES2013}. Similarly, we observed
large A$^n_N$ for $\pi^-$ ($\sim$ 5\%) compared to that for HERMES
proton data ($<$1\%)~\cite{HERMES2013}. Furthermore, we observed a large
and positive amplitude for the K$^+$ asymmetry compared to K$^-$
asymmetry on $^3$He, a similar feature observed in
$lp^{\uparrow}\rightarrow hX$ reaction on proton
target~\cite{HERMES2013}, and also the Sivers amplitude for kaons in
the SIDIS reaction at HERMES~\cite{HERMES2009}.

In summary, we have reported the first measurement of SSAs
in the inclusive hadron production reaction using unpolarized electrons on
a transversely polarized $^3$He target at $<$$p_{T}$$>=0.64$
GeV/c. Clear non-zero asymmetries were observed for charged pions and
positive kaons, showing a similar feature of flavor dependence
to that observed in the Sivers asymmetry in SIDIS, and in A$_N$ in
$pp^{\uparrow}$ collisions. Currently there are no
estimates or theoretical interpretations of these asymmetries at the
relatively low $p_T$ of 0.64 GeV/c used for this
measurement. We hope that the results presented here will
stimulate new theoretical and experimental efforts to pin-point the
exact origin of the observed SSAs. Future experiments at
Jefferson Lab~\cite{allada2012,jlab12gev}, after the 12-GeV upgrade, will
extend this measurement to higher values of $p_T$ on both proton and
$^3$He targets, and will provide precision data for future
theoretical studies. Moreover, if these non-zero asymmetries survive
at high energy kinematics then they can be used as
monitors of transverse target polarization in a fixed target
experiment, or local transverse polarization of the $^3$He beam at a
future Electron-Ion-Collider.

We acknowledge the outstanding support of the JLab Hall
A staff and the Accelerator Division in accomplishing
this experiment. This work was supported in part by the
U. S. National Science Foundation, and by Department of Energy (DOE)
contract number DE-AC05-06OR23177, under which the Jefferson Science
Associates operates the Thomas Jefferson National Accelerator
Facility.

\nocite{*}
\bibliography{references}

\end{document}

%% file: author_list.tex
% E06-010 Author list from Xiaodong Jiang



\author{K.~Allada}\email[Corresponding author: ]{kalyan@jlab.org}
\affiliation{Massachusetts Institute of Technology, Cambridge, MA 02139}
\affiliation{Thomas Jefferson National Accelerator Facility, Newport
  News, VA 23606} 
\author{Y.X.~Zhao}
\affiliation{University of Science and Technology of China, Hefei
  230026, Peoples Republic of China} 
\author{K.~Aniol}
\affiliation{California State University, Los Angeles, Los Angeles, CA 90032}
\author{J.R.M.~Annand}
\affiliation{University of Glasgow, Glasgow G12 8QQ, Scotland, United Kingdom}
\author{T.~Averett}
\affiliation{College of William and Mary, Williamsburg, VA 23187}
\author{F.~Benmokhtar}
\affiliation{Carnegie Mellon University, Pittsburgh, PA 15213}
\author{W.~Bertozzi}
\affiliation{Massachusetts Institute of Technology, Cambridge, MA 02139}
\author{P.C.~Bradshaw}
\affiliation{College of William and Mary, Williamsburg, VA 23187}
\author{P.~Bosted}
\affiliation{Thomas Jefferson National Accelerator Facility, Newport
  News, VA 23606}
\author{A.~Camsonne}
\affiliation{Thomas Jefferson National Accelerator Facility, Newport News, VA 23606}
% ...
\collaboration{The Jefferson Lab Hall A Collaboration}
\noaffiliation